# Mechanics of Cosserat media: I. non-relativistic statics

D. H. Delphenich
Kettering, OH 45440 USA

**Abstract.** The Cosserat equations for equilibrium are derived by starting from the action of the group of smooth functions with values in the Lie group of rigid spatial motions on rigid frames in Euclidian space. The method of virtual work is employed, instead of the method of action functionals, and is defined using the methods of of jet manifolds. It is shown how the restriction to fundamental forms (which define the dynamical state of the Cosserat object) that are "Euclidian" is essential to obtaining the Cosserat equations.

**Table of contents**

**1. Introduction.** In its earliest manifestations [**1**], the Cauchy stress tensor $\varepsilon_{ij}$ was considered to be symmetric in its indices. This, in turn, was a consequence of the assumption that if no external force couples were acting on a deformable body then the balance of angular momentum would imply that symmetry. However, Poisson [**2**] did comment on the meaning of asymmetry in his own research.

In Kelvin and Tate's *Treatise on Natural Philosophy* [**3**], their discussion of planar elasticity actually managed to derive some two-dimensional equations for deformation that were consistent with what the Cosserat brothers obtained later for three dimensions. Love's monumental treatise on elasticity [**4**] also discussed the role of asymmetry in the stress tensor.

Some time later, in his examination of the elastic properties of crystals that grew out of the modeling of such media by molecules, Waldemar Voigt [**5**, **6**] left open the possibility that there might be *internal* couple-stresses that acted on the molecules of the lattice. Indeed, although the atomic models that would compose crystal lattices out of atomic ions that might also impart electromagnetic polarizability to the medium would come later in history, Voigt was basically correct in his assumption. For instance, an external magnetic field will induce couple stresses in a paramagnetic solid. Such internal couple stresses would then induce an asymmetry in the Cauchy stress tensor.

Other researchers to consider the possibility of asymmetric stress tensors were Larmor [**7**], in the context of the propagation of waves in "gyrostatically loaded" materials, which related to some of the dynamical theories of the electromagnetic "ether" that preceded Maxwell's work, and Combebiac [**8**], in the context of general models for elasticity.

The Cosserat brothers, Eugène and François, first published some of their thoughts on asymmetric stress and couple-stresses as an appendix to the definitive text on mechanics by Koenigs [**9**]. They then produced their seminal work [**10**] in 1909, in which their stated objective was not so much to explore asymmetric elasticity as to explore the application to continuum mechanics of the geometry of moving frames that Darboux had defined the simplest notions of in his landmark treatise on the geometry of surfaces. Furthermore, they were going to build the variational theory of deformable bodies upon the foundations of the action of the group of Euclidian rigid motions and action functionals that were invariant under that action. Of course, they were still historically previous to the work of Cartan on the geometry of moving frames, as well as the work of Noether on the relations between symmetries of the action functional and conserved currents. Thus, to a modern reader those limitations can be somewhat discouraging, along with the daunting obesity of the mathematical notation, which involved writing out each separate coordinate in the partial derivatives and notating matrices by nine individual variables, as well as the objections of the engineers that the book also contained no applications of the abstract method to tangible examples.

Like so many seminal works (e.g., Copernicus's theory of the solar system, Mendel's paper on genetics, et al.), the book by the Cosserats remained largely unnoticed for some time, although it did get passing mention by Cartan in his own work on mechanics. The first major work to deal with it explicitly was the 1935 monograph by Sudria [**11**] in which he refined the argument of the Cosserats on action functionals that were invariant under the group of Euclidian rigid motions. Interestingly, although Noether's theorem had been published in 1916, Sudria made no mention of it in his discussion of Euclidian action functionals.

The work that followed the Cosserat work that related to its basic innovations did not always refer to that work explicitly. Mostly, it fell into two causal chains of research that eventually converged into a renewed interest in the Cosserat theory: One of these chains was concerned with the role of "asymmetric" elasticity and couple stresses, while the other one was concerned with the role played by differential geometry in the description of the yielding of elastic media into plastic deformation by the formation of dislocations and disclinations. We will deal with the latter topic in a subsequent article, since the main objective of the present article is to derive the Cosserat equations from a group-theoretic basis.

The work done on asymmetric elasticity and couple stresses seems to have begun in earnest mostly in the mid-1950's (cf. [**12-26**]). Some of the special topics that were addressed, in addition to general discussion, were stresses and strains in rods and shells [**13**], the Cosserat surface [**19**], stress concentrations in Cosserat media [**20**], stress functions in Cosserat media [**21**], and dislocations in Cosserat media [**21, 23**]. It was also observed, along the way, that anistropic stresses were also associated with liquid crystal media. One of the more definitive conferences on the subject of generalized continuum mechanics that included much that grew out of Cosserat considerations was the IUTAM symposium that was held in Stuttgart and Freudenstadt 1968 [**23**].

One of the more esoteric treatments of the Cosserat work was the 1988 book *Lie Pseudogroups and Mechanics* by Pommaret [**27**], in which he approached their theory of the mechanics of deformable bodies by using the more modern mathematical methods of Spencer cohomology, which relates to the obstructions to the integrability of over-determined systems of differential equations, such as the Lie equations that define the transformations of a Lie groupoid. A discussion of that approach will also be consigned to the subsequent article on dislocations and disclinations.

One should also confer a more recent survey on generalized continuum mechanics by Maugin [**28**] in order to get some sense of the modern thinking on the subject.

The second section of this article introduces the relevant notions that relate to the group of three-dimensional Euclidian rigid motions so the transition to differentiable functions on material objects that take their values in that group becomes a natural bridge from rigid motions to deformations. The third section then discusses the kinematics of objects in Cosserat media, in which the fundamental sequence is obtained, which relates to the integrability of deformations. The section that follows is then devoted to showing that the equations of statics for Cosserat media then follow from a principle of duality that takes the form of d'Alembert's principle of virtual work.

**2. The group of rigid motions.** Although rigid motions can be defined for any dimension of Euclidian space $E^n$, we shall concentrate on the three-dimensional case, since that is most directly relevant to the Cosserat medium.

*a. Euclidian 3-space.* Euclidian 3-space $E^3$, for us, will be a three-dimensional affine space $A^3$ that is given a Euclidian metric $g$ on its tangent bundle $T(A^3)$. Hence, we first recall some basic facts about the action of the affine group on $A^3$ and then discuss the introduction of a metric.

By definition, since $A^3$ is an affine space there is a simply transitive action $A^3 \times \mathbb{R}^3 \to A^3$, $(\mathbf{a}, x) \mapsto x + \mathbf{a}$ of the Lie group $(\mathbb{R}^3, +)$ on $A^3$ that amounts to the translation of a point to another point. By being simply transitive, we know that for any pair of points $x, y \in A^3$ there is a *unique* element $\mathbf{a} \in \mathbb{R}^3$ such that $y = x + \mathbf{a}$. Thus, one sometimes also writes $\mathbf{a} = y - x$. However, this does not generally imply that the addition of points is well-defined. In fact, one should really think of the latter difference as being a map $\mathbf{a}: A^3 \times A^3 \to \mathbb{R}^3$, $(x, y) \mapsto \mathbf{a}(x, y)$, and not an actual binary operation on $A^3$.

As a result of simple transitivity, one finds that as a differentiable manifold the Lie group $(\mathbb{R}^3, +)$ is diffeomorphic to $A^3$. However, the diffeomorphism is not unique, and, in fact, is defined by any choice of "origin" for $A^3$. Namely, if one chooses a point $O \in A^3$ then one can map every $x \in A^3$ to a unique translation vector $\mathbf{a}(O, x) \in \mathbb{R}^3$. One can also find three points $x, y, z \in A^3$ such that the corresponding translation vectors in $\mathbb{R}^3$:

$$\mathbf{e}_1 = \mathbf{a}(O, x), \quad \mathbf{e}_2 = \mathbf{a}(O, y), \quad \mathbf{e}_3 = \mathbf{a}(O, z)$$

are linearly independent, and thus define a 3-frame for either $\mathbb{R}^3$ or the tangent space $T_O(A^3)$. The ordered quadruple of points $(O, x, y, z)$ is what one calls a *reference tetrahedron* in projective geometry, and the ordered pair $(O, \mathbf{e}_i)$ is then called an *affine frame* on $A^3$.

A different choice of origin – say, $O'$ – will give a different diffeomorphism of $A^3$ with $\mathbb{R}^3$, and the three points $x, y, z$, which we shall assume are distinct from either $O$ or $O'$, define a different linear frame $\mathbf{e}'_i = \mathbf{a}(O', x_i)$ on $\mathbb{R}^3$, and thus a different affine frame $(O', \mathbf{e}'_i)$ on $A^3$. However, since the only change has been to $O$, but not $x_i$, the relationship between them is a simple translation of everything by $\pm \mathbf{d} = \mathbf{a}(O, O')$:

$$O' = O + \mathbf{d}, \quad \mathbf{e}'_i = \mathbf{e}_i - \mathbf{d}.$$

If one chooses a different set of points $x', y', z'$ to span $A^3$ with then one also has to perform an invertible linear transformation in order to convert $\mathbf{e}_i$ into $\mathbf{e}'_i = \mathbf{a}(O', x'_i)$:

$$\mathbf{e}'_i = L^{-1}(\mathbf{e}_i - \mathbf{d}).$$

Hence, if a vector $\mathbf{v} \in \mathbb{R}^3$ is expressed as $v^i \mathbf{e}_i$ or $v'^i \mathbf{e}'_i$ then since ([1]):

$$v'^i \mathbf{e}'_i = v'^i L^{-1}(\mathbf{e}_i - \mathbf{d}) = (v'^j - d^j)\tilde{L}^i_j \mathbf{e}_i = v^i \mathbf{e}_i$$

one must have:

$$v'^i = d^i + L^i_j v^j.$$

One then sees that the identity element of $A(3)$ is $(0, \delta^i_j)$ and the inverse element to $(a, L)$ is $(-\tilde{L}a, \tilde{L})$.

We define an *affine map* of $A^3$ to be a map $f: A^3 \to A^3$ that commutes with the action of the translation group, so:

$$f(x + \mathbf{d}) = f(x) + \mathbf{d}$$

for all $x \in A^3$ and all $\mathbf{d} \in \mathbb{R}^3$.

If we re-write this in terms of the left-translation operator $L_\mathbf{d}: A^3 \to A^3$, $x \mapsto x + \mathbf{d}$ as:

$$f \cdot L_\mathbf{d} = L_\mathbf{d} \cdot f \qquad \text{for all } \mathbf{d}$$

then it follows, by differentiation that:

$$df \cdot dL_\mathbf{d} = dL_\mathbf{d} \cdot df \qquad \text{for all } \mathbf{d}.$$

Now, if $\mathbf{e}_i(x)$ is a left-invariant frame field on $A^3$ then the matrix of $dL_\mathbf{d}$ with respect to $\mathbf{d}$ is $\delta^i_j$ for every $\mathbf{d}$. Hence, if $f^i_{,j}(x)$ is the matrix of $df|_x$, while $f^i_{,j}(x+\mathbf{d})$ is the corresponding matrix for $df|_{x+\mathbf{d}}$, then we must have:

$$f^i_{,j}(x+\mathbf{d}) = f^i_{,j}(x) \qquad \text{for all } \mathbf{d},$$

which simply says that the matrix $f^i_{,j}$ is a constant matrix. Thus, one can write the component representation of $y = f(x)$ in the form:

$$y^i = a^i + f^i_{,j} x^j,$$

in which the $a^i$ are integration constants, so, in fact, if $O$ goes to $0$ in $\mathbb{R}^3$ then $f(O)$ goes to $a^i$.

Hence, when one is given an affine frame $(O, \mathbf{e}_i)$, one can associate any affine map $f: A^3 \to A^3$ with a unique pair $(a^i, L^i_j) \in \mathbb{R}^3 \times GL(3)$, by setting $a^i$ equal to the components of $f(O)$ and $L^i_j$ equal to the matrix of $df$.

---

([1]) Here, and in what follows, we shall indicate the inverse of an invertible matrix by a tilde.

We then define an *affine transformation* to be an invertible affine map. If one gives the product set $\mathbb{R}^3 \times GL(3)$ the semi-direct product structure:

$$(a, L)(a', L') = (a + La', LL')$$

then the group $\mathbb{R}^3 \times_s GL(3)$ becomes isomorphic to the group $A(3)$ of affine transformations of $A^3$, and the isomorphism is defined by a choice of affine frame.

One can also define a diffeomorphism of the (trivial) bundle $GL(A^3)$ of linear frames on $A^3$ with $\mathbb{R}^3 \times GL(3)$ by taking some chosen linear frame $\mathbf{e}_i$ in the tangent space $T_O A^3$ to some chosen point $O$ – i.e., the affine frame $(O, \mathbf{e}_i)$ – and mapping it to $(0, \delta^i_j)$. Any affine frame $(x, \mathbf{f}_i)$ will go to the pair $(a^i, L^i_j)$ that makes:

$$(x, \mathbf{f}_i) = (O, \mathbf{e}_i)(a^i, L^i_j);$$

thus:

$$x = O + a^i \mathbf{e}_i, \quad \mathbf{f}_i = L^j_i \mathbf{e}_j.$$

Of course, a different choice of affine frame $(O, \mathbf{e}_i)$ will produce a different diffeomorphism.

This means that once an affine frame $(O, \mathbf{e}_i)$ is chosen, and thus, a diffeomorphism of $GL(A^3)$ with $A(3)$, one defines a *right* action $GL(A^3) \times A(3) \to GL(A^3)$ of $A(3)$ on $GL(A^3)$ by making it behave like the right-multiplication of $A(3)$ on itself:

$$(x, \mathbf{f}_i)(a^i, L^i_j) = (x + a^i \mathbf{f}_i, L^j_i \mathbf{f}_j).$$

Note that the corresponding left action would produce the essentially meaningless translation $a^i + L^i_j x$.

The definition of a metric $g$ on the tangent bundle to $A^3$ means that one introduces a scalar product $g_x(\mathbf{v}, \mathbf{w})$ on each tangent space $T_x A^3$ in a differentiable way. In order for it to be a Euclidian, it must be positive-definite, so for an orthogonal frame $\mathbf{e}_i$ in any $T_x A^3$, one will always have:

$$g_{ij}(x) = g_x(\mathbf{e}_i, \mathbf{e}_j) = \delta_{ij}. \tag{2.1}$$

In a general linear frame, the component matrix $g_{ij}(x)$ will be symmetric in the $i$ and $j$, and invertible for every $x$.

If $\theta^i$ is the reciprocal coframe to $\mathbf{e}_i$ – so $\theta^i(\mathbf{e}_j) = \delta^i_j$ – then one can express $g_x$ in the form:

$$g_x = g_{ij}(x) \, \theta^i \theta^j, \tag{2.2}$$

in which the symmetric tensor product of the 1-forms $\theta^i$ and $\theta^j$ is indicated by juxtaposing them.

*b. Finite rigid motions.* Now that we have discussed the way that the affine group $A(3)$ of $A^3$ relates to the semi-direct product $\mathbb{R}^3 \times_s GL(3)$ and the frame bundle $GL(A^3)$, and introduced a Euclidian metric on $T(E^3)$, it becomes straightforward to introduce the group $ISO(3)$ of rigid motions of $E^3$. Let $A^+(3)$ denote the subgroup of $A(3)$ that consists of the subgroup of the orientation-preserving affine transformations, which is then isomorphic to $\mathbb{R}^3 \times_s GL^+(3)$, where $GL^+(3)$ consists of all 3×3 real matrices with positive determinants. $ISO(3)$ is defined by the elements of $A^+(3)$ that also preserve the Euclidian metric $g$ on the tangent spaces to $E^3$. Thus, $M \in A^+(3)$ is a rigid motion iff:

$$g(dM|_x(\mathbf{v}), dM|_x(\mathbf{w})) = g(\mathbf{v}, \mathbf{w}) \tag{2.3}$$

for every point $x \in E^3$ and every pair of tangent vectors $\mathbf{v}$ and $\mathbf{w}$ in $T_x E^3$.

If one chooses an affine frame $(O, \mathbf{e}_i)$ on $E^3$ then the point $x \in E^3$ has the coordinates $x^i$ with respect to this frame, and the affine transformation $M$ corresponds to the element $(a^i, R^i_j)$ in $\mathbb{R}^3 \times_s GL^+(3)$, which takes $x^i$ to:

$$M(x^i) = a^i + R^i_j x^j, \tag{2.4}$$

where $R^i_j$ has a positive determinant.

The differential of the map $M: E^3 \to E^3$ takes the $a^i$ to 0 and the matrix $R^i_j$ to itself, so:

$$dM|_x = R, \tag{2.5}$$

and the defining equation (2.3) for $M$ reduces to:

$$g_{kl} R^k_i R^l_j = g_{ij}, \qquad \det[R^i_j] = 1. \tag{2.6}$$

Thus, $R^i_j$ is an orthogonal matrix with unit determinant; i.e., a three-dimensional proper rotation.

One can then show that the group $ISO(3)$ is isomorphic to the subgroup of $\mathbb{R}^3 \times_s GL^+(3)$ that is defined by $\mathbb{R}^3 \times_s SO(3)$ with the same product rule:

$$(a, R)(a', R') = (a + Ra', RR').$$

One then finds that the Lie group $ISO(3)$ is not only diffeomorphic to the product manifold $\mathbb{R}^3 \times SO(3)$, but also to the bundle $SO(E^3)$ of oriented, orthonormal frames on $E^3$. We shall return to this association later in this section.

One can then define a right action of $ISO(3)$ on $SO(E^3)$ that is the restriction of the action of $A(3)$ on $GL(E^3)$:

$$(x, \mathbf{f}_i)(a^i, R^j_i) = (x + a^i \mathbf{f}_i, \mathbf{f}_j R^j_i). \tag{2.7}$$

One should note that when one restricts the action (2.7) to the *SO*(3) subgroup that is defined by $a^i = 0$, one gets:

$$(x, \mathbf{f}_i)(0, R_i^j) = (x, \mathbf{f}_j R_i^j). \tag{2.8}$$

Thus, the *SO*(3) subgroup acts trivially on the base manifold of the principal fiber bundle *ISO*(3) → $E^3$, so we are defining the right action of *SO*(3) on the fibers as a structure group.

We should point out that the ordered set of numbers $(a^i, R_j^i)$ do not define true coordinates for *ISO*(3), since there are more components to $R_j^i$ than are necessary. Thus, what they define is an embedding $(a, R) \mapsto (a^i(a), R_j^i(R))$ of *ISO*(3) in the twelve-dimensional manifold $\mathbb{R}^3 \times M(3, 3)$ such that the image of *ISO*(3) under this embedding is defined by all $(a^i, R_j^i)$ that satisfy the defining equations (2.6). Note that this imposes no constraint on the values of $a^i$.

  *c. Infinitesimal rigid motions.* If one defines a differentiable curve $\gamma: \mathbb{R} \to ISO(3)$, $t \mapsto M(t)$ in the group of rigid motions then the velocity vectors:

$$\mathbf{v}(t) = \left. \frac{d\gamma}{dt} \right|_t$$

at each *t* can be represented by component functions with respect to some chosen oriented, orthonormal frame $(O, \mathbf{e}_i)$. That is, if $M(t) = (a^i(t), R_j^i(t))$ then:

$$\mathbf{v}(t) = \frac{da^i}{dt}(t) \frac{\partial}{\partial a^i} + \frac{dR_j^i}{dt}(t) \frac{\partial}{\partial R_j^i}.$$

This would represent the velocity, as it is seen from the point *O*.

If one wishes to describe the velocity from a co-moving frame then one must *right*-translate the point $(a^i(t), R_j^i(t))$ to $(0, \delta_j^i)$ for each *t* and the velocity components along with it. One then obtains:

$$(v^i, \omega_j^i) = \left( \frac{da^i}{dt}, \frac{dR_j^i}{dt} \right)(-\tilde{R}_k^j a^k, \tilde{R}_k^j),$$

which makes:

$$\omega_j^i = dR_k^i \tilde{R}_j^k, \qquad v^i = \frac{da^i}{dt} - \omega_j^i a^j. \tag{2.9}$$

The tangent space to the identity in *ISO*(3) can be identified with the vector space of left-invariant vector fields on *ISO*(3), and made into a Lie algebra $\mathfrak{iso}(3)$ by using the Lie bracket of vector fields. The elements of $\mathfrak{iso}(3)$ then represent the infinitesimal generators of one-parameter subgroups of rigid motions of $E^3$. As a six-dimensional Lie algebra, $\mathfrak{iso}(3)$ is isomorphic to the semi-direct sum $\mathbb{R}^3 \oplus_s \mathfrak{so}(3)$, in which the elements of $\mathbb{R}^3$ represent infinitesimal translations, and the elements of $\mathfrak{so}(3)$, which are also anti-symmetric real 3×3 matrices, represent infinitesimal rotations. The semi-direct Lie bracket makes:

$$[v + \omega, v' + \omega'] = \omega v' - \omega' v + [\omega, \omega'],$$

in which the notation $\omega v' - \omega' v$ is an abbreviation for:

$$\omega_j^i v'^j - \omega_j'^i v^j.$$

Thus, the latter expression represents the translational part of the sum, since $\mathfrak{so}(3)$ is a subalgebra, which makes $[\omega, \omega']$ another infinitesimal rotation.

The infinitesimal translations by themselves define an Abelian sub-algebra, as one sees by setting $\omega = \omega' = 0$, while the Lie bracket $[v, \omega]$ of the infinitesimal translation $v$ and the infinitesimal rotation $\omega$ is $-\omega v$. We can summarize the commutation rules for elements of $\mathfrak{iso}(3)$ as follows:

$$[v, v'] = 0, \quad [\omega, v'] = \omega v', \quad [v, \omega'] = -\omega' v, \quad [\omega, \omega'] = \omega''. \tag{2.10}$$

*d. Maurer-Cartan form on ISO*(3). Like any Lie group, *ISO*(3) admits a canonical 1-form that takes its values in the Lie algebra $\mathfrak{iso}(3)$, namely, the *Maurer-Cartan form.* It originates in the fact that any Lie group $G$ is parallelizable, so there is a global frame field on it. In fact, the way that one can show this is elementary: One simply chooses any frame $\{\mathbf{E}_a, a = 1, \ldots, N\}$ in the tangent space $T_eG$ and then left (right, resp.) translates it to the element $g \in G$ by the differential to the left (right, resp.) translation, and thus obtains a left- (right-, resp.) invariant global frame field $\{\mathbf{E}_a(g), a = 1, \ldots, N\}$ on $G$. Its reciprocal coframe field $\{\theta^a|_g, a = 1, \ldots, N\}$ is then a left- (right-, resp.) invariant coframe field on $G$.

More to the immediate point, because of the invariance of $\theta^a$, one can associate any tangent vector $\mathbf{v} \in T_gG$ with a tangent vector at $e$, namely:

$$\mathbf{v}(e) = \theta^a|_g(\mathbf{v}) \, \mathbf{E}_a = v^a(g) \, \mathbf{E}_a.$$

Since $T_eG$ can be associated with the Lie algebra $\mathfrak{g}$ of left- (right-, resp.) invariant vector fields on $G$, one can then regard the 1-form $\theta$ with:

$$\theta|_g = \theta^a|_g \otimes \mathbf{E}_a$$

as a global 1-form on $G$ with values in the Lie algebra $\mathfrak{g}$, which one calls the *Maurer-Cartan form.*

Since the $\theta^a$ define a coframe field, any 2-form can be expressed in terms of the 2-form basis $\theta^a \wedge \theta^b$, with $a < b$. In particular, the exterior derivatives of the coframe members can be expressed in the form:

$$d\theta^a = -\tfrac{1}{2} c^a_{bc} \theta^a \wedge \theta^b, \tag{2.11}$$

and from the invariance of the coframe field, the components $c^a_{bc}$ are not only constants, but, in fact, they are the *structure constants* of $\mathfrak{g}$ with respect to the frame $\mathbf{E}_a$ at $e$; that is:

$$[\mathbf{E}_b, \mathbf{E}_c] = c^a_{bc} \mathbf{E}_a. \tag{2.12}$$

Equations (2.11), which are called the *Maurer-Cartan equations* and can be concisely expressed in either forms:

$$d\theta = -\tfrac{1}{2}[\theta, \theta] \quad \text{or} \quad d\theta + \theta \wedge \theta = 0, \tag{2.13}$$

represent the statement that either the global frame field $\mathbf{E}_a$ or the global coframe field $\theta^a$ is holonomic iff the Lie algebra $\mathfrak{g}$ is Abelian.

In the case at hand, where the Lie group is *ISO*(3) and the Lie algebra is then $\mathfrak{iso}(3)$, the semi-direct product on *ISO*(3) and the corresponding semi-direct sum on $\mathfrak{iso}(3)$ allow one to decompose the tangent bundle $T(ISO(3)) = T(\mathbb{R}^3 \times SO(3))$ into a Whitney sum $T(\mathbb{R}^3) \oplus T(SO(3))$, so the tangent space to the identity decomposes into $\mathbb{R}^3 \oplus \mathfrak{so}(3)$, as a vector space. Similarly, if one chooses $\mathbf{E}_a$ to be an adapted frame for this decomposition of $T_e ISO(3)$ – so, by definition, three of the frame members (say $a = 1, 2, 3$) span the $\mathbb{R}^3$ subspace and the other three (say, $a = 4, 5, 6$) span the $\mathfrak{so}(3)$ subspace – then the coframe field $\theta^a$ will also be adapted to the decomposition $\Lambda^1(ISO(3)) = \Lambda^1(\mathbb{R}^3) \oplus \Lambda^1(SO(3))$. Thus, $\theta^1, \theta^2, \theta^3$ will take their values in the $\mathbb{R}^3$ subalgebra of $\mathfrak{iso}(3)$ and $\theta^4, \theta^5, \theta^6$ will take their values in the $\mathfrak{so}(3)$ subalgebra. For simplicity, we change the notation for the last three to $\theta^i$, $i = 1, 2, 3$, respectively, while $a = 1, 2, 3$, now.

One can then subdivide the Maurer-Cartan equations into equations for the exterior derivatives of the two sets of 1-forms thus defined:

$$d\theta^a = -\tfrac{1}{2} c^a_{bc} \theta^b \wedge \theta^c - c^a_{bi} \theta^b \wedge \theta^i - \tfrac{1}{2} c^a_{ij} \theta^i \wedge \theta^j, \tag{2.14}$$
$$d\theta^i = -\tfrac{1}{2} c^i_{bc} \theta^b \wedge \theta^c - c^i_{bi} \theta^b \wedge \theta^i - \tfrac{1}{2} c^i_{jk} \theta^j \wedge \theta^k. \tag{2.15}$$

From the rules (2.10), one sees that since $\mathbb{R}^3$ is Abelian, $c^*_{bc} = 0$ in any case, while the fact that $[\mathbf{E}_b, \mathbf{E}_i] = -\varepsilon_{bia} \mathbf{E}_a$ is in $\mathbb{R}^3$ makes $c^a_{bi} = -\varepsilon_{bia}$ and $c^j_{bi} = 0$. Finally, the fact that $\mathfrak{so}(3)$ is a subalgebra means that $[\mathbf{E}_j, \mathbf{E}_k] = \varepsilon_{jki} \mathbf{E}_i$, which makes $c^a_{jk} = 0$ and $c^i_{jk} = \varepsilon_{jki}$. Inserting all of these specifications into (2.14) and (2.15) gives:

$$d\theta^a = \tfrac{1}{2}\varepsilon_{bia}\theta^b \wedge \theta^i, \qquad d\theta^i = -\tfrac{1}{2}\varepsilon_{jki}\theta^j \wedge \theta^k. \tag{2.16}$$

The exterior differential system $\theta^a = 0$ defines the sub-bundle of $\mathfrak{so}(3)$ subspaces in $T(ISO(3))$, while $\theta^i = 0$ defines the sub-bundle of $\mathbb{R}^3$ subspaces. One sees from (2.16) that the latter sub-bundle is in involution, and thus integrable, while the former one is not. That is, the $\mathfrak{so}(3)$ subspaces have become essentially "twisted" by the semi-direct product structure into a non-integrable sub-bundle, despite the fact that $ISO(3)$ is a product manifold. Due to the product structure, there should be an adapted coframe field on $ISO(3)$ such that the $\mathbb{R}^3$ part of $T(ISO(3))$ is integrable into the $\mathbb{R}^3$ subgroup of $ISO(3)$ and the $\mathfrak{so}(3)$ part of $T(ISO(3))$ is integrable into the $SO(3)$ subgroup. Thus, the adapted coframe field would not be the Maurer-Cartan 1-form.

Under the identification of $ISO(3)$ with $SO(E^3)$ by some choice of diffeomorphism, one sees that the Maurer-Cartan 1-form $\theta$ on $ISO(3)$ corresponds to a 1-form on $SO(E^3)$ with values in $\mathfrak{iso}(3)$ that we shall also denote by $\theta$ and refer to as the Maurer-Cartan form on $SO(E^3)$. Although it is tempting to think of $\theta$, when adapted to the decomposition of $T(SO(E^3))$ into $T(E^3) \oplus T(SO(3))$, as the union of the canonical 1-forms $\theta^a$ on $SO(E^3)$ with an $\mathfrak{so}(3)$ connection 1-form $\theta^i$, one must remember that we are more immediately concerned with a right action of $ISO(3)$ on $SO(E^3)$ that does not behave like the right action of $SO(3)$ on that manifold when one restricts to that subgroup.

*e. Actions of rigid motions on manifolds.* A Lie group $G$ is said to *act differentiably (on the left)* on the differentiable manifold $M$ iff there is a differentiable map $G \times M \to M$, $(g, x) \mapsto gx$ such that:
1. $ex = x$, where $e$ is the identity element of $G$.
2. $g(g'x) = (gg')x$.

As a consequence of this, one also must have:
3. If $y = gx$ then $x = g^{-1}y$.

One can define a right action analogously. Generally, the effect of right action is distinct from the effect of left action. The necessity for considering both types of actions often relates the fact that square matrices act on column vectors from the left and row vectors from the right.

For a given $g \in G$ one can then define a diffeomorphism $L_g : M \to M$, $x \mapsto L_g x = gx$ that one calls *left translation by g*. If Diff($M$) represents the group of diffeomorphisms of $M$ to itself then the resulting map $G \to$ Diff($M$), $g \mapsto L_g$ is then a group homomorphism. For a right action, one similarly defines right translation and its representation as a subgroup of Diff($M$).

An important example of a group action is the *representation* of a Lie group *G* in a vector space *V*, which can be defined to be either a homomorphism $\sigma: G \to GL(V)$ or a linear (left or right, resp.) action of *G* on *V* that takes any (*g*, *v*) to $\sigma(g)v$ or $v\sigma(g)$, respectively, depending upon which is more natural. When *G* is represented by a subgroup of *GL*(*n*) – i.e., by invertible $n \times n$ matrices – then there is a natural left linear action of *G* on $\mathbb{R}^n$ and a right action on $\mathbb{R}^{n*}$ that one simply calls the *defining representation*.

Given two points *x* and *y* of *M* there can generally be no elements of *G*, a unique element of *G*, or more than one element of *G* that makes *y* = *gx*. When there is always at least one, one calls the action *transitive*, and when there is exactly one, one calls it *simply transitive*.

By differentiating the group action at the identity of *G*, one gets an infinitesimal action of the Lie algebra $\mathfrak{g}$ on *M* to produce tangent vectors to *M*. In particular, for a given $\mathfrak{a} \in \mathfrak{g}$ there is a unique global vector field $\hat{\mathfrak{a}}$ on *M* that one calls the *fundamental vector field* that is associated with $\mathfrak{a}$ and is defined by:

$$\hat{\mathfrak{a}}(x) = \frac{d}{ds}\bigg|_{s=0} \exp(\mathfrak{a}s)x, \qquad (2.17)$$

in which $\exp(\mathfrak{a}s)$ refers the exponential map that takes elements of $\mathfrak{g} = T_e G$ to elements of *G*. When *G* is a matrix Lie group, one has, in fact:

$$\exp(\mathfrak{a}) = \sum_{n=0}^{\infty} \frac{1}{n!} \mathfrak{a}^n. \qquad (2.18)$$

One then has a linear map $\mathfrak{g} \to \mathfrak{X}(M)$, $\mathfrak{a} \mapsto \hat{\mathfrak{a}}$ that proves to be a Lie algebra homomorphism. (Here, $\mathfrak{X}(M)$ refers to the infinite-dimensional Lie algebra of vector fields on *M*.)

   *f. The dual of the Lie algebra of rigid motions.* The dual of the vector space $\mathfrak{iso}(3)$ consists of all linear functionals on $\mathfrak{iso}(3)$, and will be denoted by $\mathfrak{iso}(3)^*$. However, the vector space $\mathfrak{iso}(3)^*$ is not generally given a Lie bracket. Just as the vector space $\mathfrak{iso}(3)$ is isomorphic to $\mathbb{R}^3 \oplus \mathfrak{so}(3)$, similarly, the vector space $\mathfrak{iso}(3)^*$ is isomorphic to $\mathbb{R}^{3*} \oplus \mathfrak{so}(3)^*$. That is, an element $P \in \mathfrak{iso}(3)^*$ can be represent by a sum:

$$P = p + L,$$

in which $p \in \mathbb{R}^{3*}$ and $L \in \mathfrak{so}(3)^*$.

Although we are using notation that suggests linear and angular momentum as the physical interpretation for an element of $\mathfrak{iso}(3)^*$, actually, one can just as well interpret any element of that vector space as a force combined with a force moment. Such a concept goes back to the Nineteenth Century discussions on the projective geometry of statics and mechanics by Poinsot, Chasles, Plücker, Klein, Ball, and Study that referred to such objects as "torsors," "dynamen," or "wrenches," depending upon the language of the researcher.

If $\xi = v + \omega$ is an element of $\mathfrak{iso}(3)$ then the evaluation of $P$ on $\xi$ is given by:

$$P(\xi) = p(v) + L(\omega).$$

If the element $v + \omega \in \mathfrak{iso}(3)$ represents a linear and angular velocity, while the dual object $p + L \in \mathfrak{iso}(3)^*$ represents a linear and angular momentum then the real number $P(\xi)$ represents a total kinetic energy. However, if $v + \omega$ represents an infinitesimal (virtual) displacement (both linear and angular), while $p + L$ represents a force and a force couple, then that number would represent a virtual work done by the force over that displacement.

There is a dual to the Maurer-Cartan 1-form $\theta$ on $ISO(3)$ that amounts to a vector field $\mathbf{E}$ on $ISO(3)$ with coefficients in $\mathfrak{iso}(3)^*$ that is defined by:

$$\mathbf{E} = \mathbf{E}_i \otimes \theta^i |_{(0, I)} + \mathbf{E}_a \otimes \theta^a |_{(0, I)} . \qquad (2.19)$$

When it is evaluated on an arbitrary vector field $\mathbf{v}$ on $ISO(3)$ the result is a left-invariant vector field on $ISO(3)$ that agrees with $\mathbf{v}(0, I)$:

$$\mathbf{E}(\mathbf{v}) = v^i(0, I) \, \mathbf{E}_i + v^a(0, I) \, \mathbf{E}_a . \qquad (2.20)$$

The equations that are dual to the Maurer-Cartan equations are simply the structure equations (2.12).

*g. The bundle of oriented, orthonormal frames on $E^3$.* As mentioned above, as a manifold, the Lie group $ISO(3)$ is diffeomorphic to the manifold $SO(E^3)$, which is the total space of the (trivial) bundle of oriented, orthonormal frames on $E^3$. We shall now focus somewhat more attention on this latter manifold.

The manifold $SO(E^3)$ is diffeomorphic to the product manifold $\mathbb{R}^3 \times SO(3)$, so it is six-dimensional, connected, but not simply-connected, and locally-compact, but not compact. Indeed, its homotopy type is that of $SO(3)$, which is diffeomorphic to $\mathbb{R}P^3$, whose homotopy groups differ from those of $S^3$ only in dimension one, in which $\pi^1(\mathbb{R}P^3) = \mathbb{Z}_2$.

Since $SO(E^3)$ is diffeomorphic to a Lie group, it is parallelizable. Therefore, global frame fields and coframe fields will exist on it. However, since it is not diffeomorphic to a vector space, but only the $\mathbb{R}^3$ factor in the product, it will not admit a global coordinate system.

As we have mentioned before, since it is the matrix representations of Euclidian rotations that are of most interest to us, whenever we associate a rigid frame $(x, \mathbf{f}_i)$ in $SO(E^3)$ with numbers $(x^i, f^i_j)$ that represent the coordinates of $(x, \mathbf{f}_i)$ with respect to some choice of reference frame $(O, \mathbf{e}_i)$, so:

$$(x, \mathbf{f}_i) = (O, \mathbf{e}_i)(x^i, f^i_j) = (O + \mathbf{e}_i x^i, \mathbf{e}_i f^i_j), \qquad (2.21)$$

we are actually defining an embedding of the six-dimensional manifold $SO(E^3)$ in the twelve-dimensional vector space $\mathbb{R}^3 \times M(3, 3)$, whose coordinates can be described by ordered 12-tuples of numbers of the form $(a^i, M^i_j)$. The image of $ISO(3)$ under the embedding then consists of all points that satisfy the defining algebraic equations (2.6) for rigid motions.

One can also regard the basic equation (2.21) as the definition of a global right action of $ISO(3)$, when represented by matrices, on the manifold $SO(E^3)$. That is, the right action of the group element $(x^i, f^i_j) \in ISO(3)$ on the rigid frame $(O, \mathbf{e}_i) \in SO(E^3)$ gives the rigid frame $(x, \mathbf{f}_i)$. However, this right action on the *bundle $SO(E^3)$* is not the action of a structure group on a principal fiber bundle, since the structure group of $SO(E^3)$ is $SO(3)$, not $ISO(3)$ and the action of a structure group on a principal bundle is "vertical," in the sense that it takes elements of a given fiber to other elements of the *same* fiber, while this action takes frames at one point to frames at a translated point, in general.

This global right action of $ISO(3)$ on $SO(E^3)$ also allows a notion of right-invariant vector fields and 1-forms on $SO(E^3)$, and thus, right-invariant frame fields and coframe fields. The components of a right-invariant vector field or 1-form with respect to a right-invariant frame field or coframe field, resp., will then be constants. Similarly, there will be an equivalent to the Maurer-Cartan equations for any global coframe field $\theta^a$, $a = 1, \ldots, 6$ on $SO(E^3)$, namely:

$$d\theta^a = -\tfrac{1}{2} c^a_{bc}(x, \mathbf{f}_i) \theta^b \wedge \theta^c, \qquad (2.22)$$

in which the functions $c^a_{bc}(x, \mathbf{f}_i)$ are the *structure functions* of the coframe field $\theta^a$; if $\theta^a$ is right-invariant then these functions will be constant.

If the coframe field $\theta^a$ is reciprocal to the frame field $\mathbf{e}_a$ then the structure equations that are dual to (2.22) take the form:

$$[\mathbf{e}_b, \mathbf{e}_c] = c^a_{bc}(x, \mathbf{f}_i) \mathbf{e}_a, \qquad (2.23)$$

which generalize the definition of the structure constants for a Lie algebra.

**3. Kinematics of Cosserat media.** Now that we have made the preparatory definitions in the language of the group $ISO(3)$ of rigid motions and the bundle $SO(E^3)$ of oriented, orthonormal frames on $E^3$, we can specialize them to the physical application of Cosserat media.

We define a *Cosserat medium* – or *Cosserat object* – to be an embedding $f: \mathcal{O} \to SO(E^3)$, $\rho^a \mapsto (x(\rho), \mathbf{e}_i(\rho))$, in which $\mathcal{O} \subset \mathbb{R}^p$ is a subset of a $p$-dimensional parameter space. Generally, it will be a closed subset that represents a manifold with boundary $\partial \mathcal{O}$. In conventional continuum mechanics, the parameters $\rho^a$, $a = 1, \ldots, p$ are sometimes referred to as "material coordinates." Typically, $p$ is 1, 2, or 3 for non-relativistic statics problems, which would correspond to the deformation of strings, surfaces, and solids.

The way that we shall impart motion or deformation to a Cosserat object is by extending the right action of $ISO(3)$ on $SO(E^3)$ to a right action of $C^\infty(\mathcal{O}, ISO(3))$ on $C^\infty(\mathcal{O}, SO(E^3))$, where $C^\infty(\mathcal{O}, ISO(3))$ is the group of all smooth maps from $\mathcal{O}$ to $ISO(3)$, with point-wise multiplication, and $C^\infty(\mathcal{O}, SO(E^3))$ is the set of all smooth embeddings. The action is then:

$$(x(\rho), \mathbf{e}_i(\rho)) = (x_0(\rho), \mathbf{e}_{0i}(\rho))(\tilde{a}^i(\rho), \tilde{R}^i_j(\rho)), \tag{3.1}$$

with:

$$x(\rho) = x_0(\rho) + \tilde{a}^i(\rho)\mathbf{e}_{0i}(\rho), \tag{3.2}$$

$$\mathbf{e}_i(\rho) = \mathbf{e}_{0j}(\rho)\tilde{R}^j_i(\rho). \tag{3.3}$$

Actually, the group $C^\infty(\mathcal{O}, ISO(3))$ and the set $C^\infty(\mathcal{O}, SO(E^3))$ are not as practical to deal with as the manifolds $J^k(\mathcal{O}, ISO(3))$ and $J^k(\mathcal{O}, SO(E^3))$ of $k$-jets of $C^k$ maps from $\mathcal{O}$ to $ISO(3)$ and $SO(E^3)$. Namely, in order to turn the group $C^\infty(\mathcal{O}, ISO(3))$ into an infinite-dimensional Lie group and to turn the set $C^\infty(\mathcal{O}, ISO(E^3))$ into an infinite-dimensional manifold requires considerable analytical overhead, while the manifolds $J^k(\mathcal{O}, ISO(3))$ and $J^k(\mathcal{O}, SO(E^3))$ are finite-dimensional, as well as being immediately applicable to the problems of mechanics. Thus, we shall introduce the elementary language and machinery of jets in a later subsection.

In order to abbreviate the notation, when we are dealing with spaces of smooth – or at least "sufficiently differentiable" – maps from $\mathcal{O}$ to other spaces, we shall use only the range manifold in the notation. Thus, for instance, $C^\infty(\mathcal{O}, ISO(E^3))$ abbreviates to just $ISO(E^3)$ and differential $k$-forms on $\mathcal{O}$ will be elements of $\Lambda^k \mathcal{O}$, not $C^\infty(\mathcal{O}, \Lambda^k \mathcal{O})$. Although this notation can lead to confusion, one notes that generally differential operators are only meaningful on functions and sections, so in most cases, the notation would be otherwise meaningless.

*a. Kinematical state of a Cosserat object.* We now drop the explicit reference to the parameter $\rho$ for the sake of brevity. We suppose that a motion or deformation of an object $(x, \mathbf{e}_i)$ is the result of the action of an element $(\tilde{a}^i, \tilde{R}^i_j)$ on an initial object $(x_0, \mathbf{e}_{i0})$, which we write as a pair of equations:

$$x = x_0 + \tilde{a}^i \mathbf{e}_{i0}, \qquad \mathbf{e}_i = \mathbf{e}_{j0}\tilde{R}^j_i, \tag{3.4}$$

which is an abbreviation for (3.2), (3.3).

One sees that one advantage of the Cosserat approach is that the main difference between a deformation and a motion is whether one of the parameters is identified as "temporal," while the other ones are identified as "spatial." However, in this article, we shall deal with statics, so we shall only refer to deformations.

Corresponding to the right action on frames, one has a left action on components:

$$x^i = a^i + R^i_j x^j_0, \qquad f^i_j = R^i_k f^k_{0j}. \tag{3.5}$$

We shall call this the *displacement* of the initial object. It relationship to the usual displacement of conventional continuum mechanics is in:

$$u^i(x_0) = x^i - x^i_0 = a^i + (R^i_j - \delta^i_j) x^j_0. \tag{3.6}$$

In Schaefer [**22**], the expression $R^i_j - \delta^i_j$ is assumed to be infinitesimal and gets (implicitly) replaced with the anti-symmetric matrix $\varphi^i_j$, and then with the 3-vector $\varphi^j$ that makes:

$$\varphi_{ij} = \mathrm{ad}(\varphi^j) = \varepsilon_{ijk}\, \varphi^k, \tag{3.7}$$

in which ad refers to the "adjoint" representation of the Lie algebra $\mathfrak{so}(3)$ in $\mathfrak{gl}(3)$, which takes the infinitesimal rotation vector $\boldsymbol{\omega} = (\omega^1, \omega^2, \omega^3)$ to the infinitesimal rotation matrix $\mathrm{ad}(\boldsymbol{\omega})$, which is the matrix of the linear map that takes every vector $\mathbf{v} \in \mathbb{R}^3$ to $\boldsymbol{\omega} \times \mathbf{v}$.

Shaefer's notation for the displacement then amounts to:

$$\chi(\rho) = \begin{bmatrix} \varphi^i \\ u^i - \varepsilon^{ijk} \varphi_j x_{0k} \end{bmatrix}. \tag{3.8}$$

For us, it shall be:

$$\chi(\rho) = \begin{bmatrix} a^i \\ R^i_j \end{bmatrix} = \begin{bmatrix} x^i - R^i_j x^j_0 \\ R^i_j \end{bmatrix}, \tag{3.9}$$

in order to be consistent with the transpose of $(a, R)$. Upon replacing $x^i$ with $x^i_0 + u^i$, and $R^i_j$ with $\delta^i_j + \varphi^i_j$, one sees that the relationship between our notation and his is simply:

$$\begin{bmatrix} a^i \\ R^i_j \end{bmatrix} = \begin{bmatrix} u^i - \varphi^i_j x^j_0 \\ \delta^i_j + \varphi^i_j \end{bmatrix}, \tag{3.10}$$

which is essentially a permutation of the components, except for the addition of $\delta^i_j$, which will vanish upon differentiation.

When we get to variations of the kinematical state, we shall see that, in effect, we are transferring the burden of accounting for the deformation from the deformed state *f* itself to the displacement $\chi$, so the equations of equilibrium and dynamics eventually become equations on functions that take their values in the group of rigid motions.

Differentiating $\chi$ with respect to $\rho$ then gives the *deformation* of the object:

$$d\chi = \begin{bmatrix} da^i \\ dR^i_j \end{bmatrix}. \tag{3.11}$$

This is the way that the deformation is described from the initial frame $(O, \mathbf{e}_i)$, which represents the *Lagrangian* picture. With respect to the deformed object $f$, which is called the *Eulerian picture*, one must right-translate $\chi$ back to $(0, \delta^i_j)$ by way of the inverse of the displacement, namely:

$$\chi^{-1} = \begin{bmatrix} -\tilde{R}^i_j a^i \\ \tilde{R}^i_j \end{bmatrix}, \tag{3.12}$$

which makes:

$$E \equiv d\chi \chi^{-1} = \begin{bmatrix} da^i - dR^i_k \tilde{R}^k_j a^j \\ dR^i_k \tilde{R}^k_j \end{bmatrix} = \begin{bmatrix} \xi^i \\ \omega^i_j \end{bmatrix}, \tag{3.13}$$

in which we have defined the generalized velocities:

$$\omega^i_j = dR^i_k R^k_j, \quad \xi^i = da^i - \omega^i_j a^j, \tag{3.14}$$

which comes about much like the angular velocity of a rotating frame, except that the derivatives are partial derivatives with respect to spatial parameters, now.

One finds by simple calculations that:

$$\begin{bmatrix} \xi^i \\ \omega^i_j \end{bmatrix} = \begin{bmatrix} da^i - \omega^i_j a^j \\ \omega^i_j \end{bmatrix} = \begin{bmatrix} dx^i - \omega^i_j x^j \\ \omega^i_j \end{bmatrix} = \begin{bmatrix} du^i - \omega^i_j x^j \\ \omega^i_j \end{bmatrix}. \tag{3.15}$$

It is important to note that the second form of the expressions involves only variables that take their values in the group of rigid motions and its Lie algebra.

Thus, if we denote the translational part of $E$ by the set of three 1-forms on $\mathcal{O}$:

$$E^i = E^i_a d\rho^a = (u^i_{,a} - \omega^i_{ja} x^j) d\rho^a \tag{3.16}$$

then we can see how this definition of deformation relates to the conventional definitions (cf., e.g., [**4, 5, 15, 25**]) of infinitesimal strain and rotation, which are defined by:

$$e_{ij} = u_{i,j} + u_{j,i}, \qquad \theta_{ij} = u_{i,j} - u_{j,i}, \tag{3.17}$$

respectively. Namely, if we consider a three-dimensional object $\mathcal{O}$ that represents a subset of $E^3$, so the embedding is the inclusion and the parameters $\rho^\alpha$ become the initial coordinates $x_0^i$, then we can see that the components of the 1-forms $E^i$, when the upper one is lowered using the Euclidian metric, are:

$$E_{ij} = u_{i,j} - \omega_{kj} x^k. \tag{3.18}$$

Its symmetric part is:

$$E_{(ij)} = u_{(i,j)} - \omega_{(i|k|j)} x^k = \tfrac{1}{2} e_{ij} - \omega_{(i|k|j)} x^k, \tag{3.19}$$

and its anti-symmetric part is:

$$E_{[ij]} = u_{[i,j]} - \omega_{[i|k|j]} x^k = \tfrac{1}{2} \theta_{ij} - \omega_{[i|k|j]} x^k, \tag{3.20}$$

so one has:

$$E_{ij} = \tfrac{1}{2}(e_{ij} + \theta_{ij}) - \omega_{kj} x^k. \tag{3.21}$$

Thus, one sees that generally the infinitesimal rotation 1-form $\omega_j^i$ is independent of the infinitesimal rotation $\theta_{ij}$ that comes from the displacement. This situation is analogous to the difference between orbital angular momentum and intrinsic angular momentum, or "spin." We shall refer to the matrix of 1-forms $\omega_j^i$ as the *infinitesimal couple-strain*, which will be dual to the couple-stress that we shall introduce later.

One already sees that the presence of a derivative operator in the definition of deformation means that there will be more than one element of $C^\infty(\mathcal{O}, ISO(3))$ that goes to 0. In fact, the kernel of the operator $d: ISO(3) \to \Lambda^1 \otimes T(ISO(3))$, $(a, R) \mapsto (da, dR)$ consists of all constant smooth functions from the object $\mathcal{O}$ to the group of rigid motions in the space $E^3$, which then means that rigid motions of the object produce no deformation. This constraint will prove crucial in the derivation of the Cosserat equations.

We shall now introduce a notation that suggests exterior covariant differentials, although one should be wary of giving in to that tendency, since connections have more to do with actions of groups on fibers of bundles that do not affect the points of the base manifold, while the action that we have defined on $SO(E^3)$ does not "cover the identity" the way that the action of $SO(3)$ on the frames as a structure group does. In particular, it moves an oriented, orthonormal frame $\mathbf{e}_i$ at a point $x$ to another such frame at the point $x + a^i \mathbf{e}_i$. Similarly, the 1-form on $\mathcal{O}$ with values in the Lie algebra $\mathfrak{iso}(3)$ that takes the form of $(\xi, \omega) = (da - \omega a, \omega)$ should not be regarded as an $ISO(3)$ connection 1-form in the precise sense, since a connection would be an infinitesimal generator of a one-parameter family of rigid motions in the affine tangent spaces to $E^3$, not rigid motions of the frames of $SO(E^3)$. Furthermore, unlike the set of 1-forms $\theta^i$, $i = 1, 2, 3$ that locally represent the canonical 1-form on $SO(E^3)$, the set of 1-forms $\xi^i = da^i - \omega a^i$ on $E^3$ does not have to define a coframe field. As generalized velocities, not only do they not have to be linearly independent, they can even be zero.

However, one can think of $E$ as the pull-back to $\mathcal{O}$ of the Maurer-Cartan 1-forms on $ISO(3)$ by the embedding $g: \mathcal{O} \to ISO(3)$, or similarly, the pull-back of those 1-forms on $SO(E^3)$ by the embedding $f: \mathcal{O} \to SO(E^3)$.

We define the map $\nabla: ISO(3) \to \Lambda^1 \mathcal{O} \otimes \mathfrak{iso}(3)$, $\chi \mapsto \nabla \chi$ by:

$$\nabla \chi = d\chi\, \chi^{-1}, \tag{3.22}$$

as above, so:

$$\omega \equiv \nabla R = dR\, R^{-1}, \qquad \xi \equiv \nabla a = da - \omega a. \tag{3.23}$$

If one takes the exterior derivative of $\nabla \chi$ then one finds that:

$$d_\wedge \omega = -dR \wedge dR^{-1} = -dR\, R^{-1} \wedge R\, dR^{-1} = \omega \wedge \omega,$$
$$d_\wedge \xi = d_\wedge(da - \omega a) = -d_\wedge \omega\, a + \omega \wedge da = -\omega \wedge \omega a + \omega \wedge \xi - \omega \wedge \omega a = \omega \wedge \xi,$$

in which we have used the fact that since $R^{-1} R = I$, one must have:

$$\omega = dR\, R^{-1} = -R\, dR^{-1}. \tag{3.24}$$

Thus, one obtains a compatibility – or integrability – condition on a given $E = (\xi, \omega) \in \Lambda^1 \mathcal{O} \otimes \mathfrak{iso}(3)$ to be of the form $\nabla \chi$ for some displacement $\chi$, namely, one must have:

$$0 = d_\wedge \xi - \omega \wedge \xi, \qquad 0 = d_\wedge \omega - \omega \wedge \omega. \tag{3.25}$$

If we then extend the map $\nabla$ to a map $\nabla: \Lambda^1 \mathcal{O} \otimes \mathfrak{iso}(3) \to \Lambda^2 \mathcal{O} \otimes \mathfrak{iso}(3)$ by setting:

$$\nabla_\wedge E = \begin{bmatrix} \Theta \\ \Omega \end{bmatrix} = \begin{bmatrix} d_\wedge \xi - \omega \wedge \xi \\ d_\wedge \omega - \omega \wedge \omega \end{bmatrix} \tag{3.26}$$

then the compatibility condition is simply that $E = \nabla \chi$ only if:

$$\nabla_\wedge E = 0. \tag{3.27}$$

We shall return to the question of whether this condition is also sufficient in a later paper.

Although it is tempting to refer to the 2-forms $\Theta^i$ as the torsion of the connection $(\xi, \omega)$, this is not precisely true, since we said that $(\xi, \omega)$ is not really a connection, to begin with, and furthermore, the 1-forms $\xi^i$ do not have to define a coframe field; indeed, some of them might be zero. Similarly, the 2-forms $\Omega^i_j$ are closely analogous to curvature 2-forms, except that, once again, $(\xi, \omega)$ is not a connection. Thus, we shall only refer to the 2-form $\Theta$ with values in $\mathfrak{iso}(3)$ as the *dislocation* of the deformation $E$, and the 2-form $\Omega$ with values in $\mathfrak{iso}(3)$ will be its *disclination*. Thus, a deformation with non-vanishing dislocation or disclination admits no global displacement.

If one expresses the 1-forms involved in terms of their components with respect to the 1-forms $d\rho^a$:

$$\xi^i = \xi^i_a \, d\rho^a, \qquad \omega^i_j = \omega^i_{ja} \, d\rho^a \qquad (3.28)$$

then the local form of the compatibility equations for $(\xi, \omega)$ is:

$$\begin{cases} 0 = \dfrac{\partial \xi^i_b}{\partial \rho^a} - \dfrac{\partial \xi^i_a}{\partial \rho^b} - \omega^i_{ja}\xi^j_b + \omega^i_{jb}\xi^j_a, \\ 0 = \dfrac{\partial \omega^i_{jb}}{\partial \rho^a} - \dfrac{\partial \omega^i_{ja}}{\partial \rho^b} - \omega^i_{ka}\omega^k_{jb} + \omega^i_{kb}\omega^k_{ja}. \end{cases} \qquad (3.29)$$

One can find analogues of these equations in not only the Cosserat work (e.g., [**10**], Chap. IV, sec. **50**), but also the classic text on theoretical kinematics by Koenigs [**9**].

One sees in the latter form of the compatibility conditions that one is demanding that the anti-symmetric parts of the second-rank tensors on $\mathcal{O}$:

$$\nabla \xi^i_a = \left( \dfrac{\partial \xi^i_b}{\partial \rho^a} - \omega^i_{ja}\xi^i_b \right) d\rho^a \otimes d\rho^b, \qquad (3.30)$$

$$\nabla \omega^i_{ja} = \left( \dfrac{\partial \omega^i_{jb}}{\partial \rho^a} - \omega^i_{ka}\omega^k_{jb} \right) d\rho^a \otimes d\rho^b, \qquad (3.31)$$

must vanish.

It is essential to understand that this way of characterizing the compatibility of a deformation by means of a first-order differential operator has a distinct advantage over the conventional St.-Venant compatibility conditions for infinitesimal strains [**4, 15, 25**], which involve a second-order differential operator and amount to the vanishing of the curvature of the deformed metric.

One can go one step further and look for compatibility conditions that would make a given 2-form $(\Theta, \Omega)$ with values in $\mathfrak{iso}(3)$ take the form of $\nabla \wedge E$ by first taking the exterior derivatives of both sides of the defining equations:

$$\Theta = d_\wedge \xi - \omega \wedge \xi, \qquad \Omega = d_\wedge \omega - \omega \wedge \omega, \qquad (3.32)$$

which gives:

$$\begin{aligned} d_\wedge \Theta &= - d_\wedge \omega \wedge \xi + \omega \wedge d_\wedge \xi = -(\Omega + \omega \wedge \omega) \wedge \xi + \omega \wedge (\Theta + \omega \wedge \xi) \\ &= - \Omega \wedge U + \omega \wedge \Theta. \end{aligned}$$

$$\begin{aligned} d_\wedge \Omega &= - d_\wedge \omega \wedge \omega + \omega \wedge d_\wedge \omega = -(\Omega + \omega \wedge \omega) \wedge \omega + \omega \wedge (\Omega + \omega \wedge \omega) \\ &= - \Omega \wedge \omega + \omega \wedge \Omega - (\omega \wedge \omega) \wedge \omega + \omega \wedge (\omega \wedge \omega). \end{aligned}$$

The last two terms, which should really be written:

$$-\tfrac{1}{2}([[\omega,\omega],\omega]-[\omega,[\omega,\omega]]),$$

will vanish from the Jacobi identity, and if one introduces the notation:

$$\nabla_\wedge \begin{bmatrix} \Theta \\ \Omega \end{bmatrix} = \begin{bmatrix} d_\wedge \Theta - \omega \wedge \Theta + \Omega \wedge \xi \\ d_\wedge \Omega - \omega \wedge \Omega + \Omega \wedge \omega \end{bmatrix} \quad (3.33)$$

then the compatibility condition for $(\Omega, \Theta)$ to be of the form $\nabla_\wedge E$ for some $E$ can be written:

$$\nabla_\wedge \begin{bmatrix} \Theta \\ \Omega \end{bmatrix} = 0. \quad (3.34)$$

Once again, although it looks as if we are dealing with the Bianchi identities, in effect, these identities are distinct from them. We shall simply refer to the 3-form $(\nabla_\wedge \Omega, \nabla_\wedge \Theta)$ with values in $\mathfrak{iso}(3)$ as the *incompatibility* in $(\Omega, \Theta)$.

Thus we have defined a sequence of first-order differential operators ([1]), which we refer to as the *fundamental sequence* for Cosserat kinematics:

$$ISO(3) \xrightarrow{\nabla_\wedge} \Lambda^1 \mathcal{O} \otimes \mathfrak{iso}(3) \xrightarrow{\nabla_\wedge} \Lambda^2 \mathcal{O} \otimes \mathfrak{iso}(3) \xrightarrow{\nabla_\wedge} \Lambda^3 \mathcal{O} \otimes \mathfrak{iso}(3)$$

that is *cohomological*, in the sense that one has:

$$\nabla_\wedge^2 = 0, \quad (3.35)$$

at each step. Whether it is also exact at each step will be the subject of a later paper.

*b. Kinematical states and jets.* We are now in a position to define the kinematical state of an object $\mathcal{O}$ that is embedded in a Cosserat medium $SO(E^3)$ to be a section of the "source projection" $J^1(\mathcal{O}, SO(E^3)) \to \mathcal{O}$, so we first define what that would mean.

The manifold $J^1(\mathcal{O}, SO(E^3))$ consists of all *1-jets* ([2]) of differentiable embeddings $f: \mathcal{O} \to SO(E^3)$, $\rho \mapsto (x^i(\rho), \mathbf{e}_i(\rho))$, where the 1-jet $j_\rho^1 f$ of $f$ at $\rho \in \mathcal{O}$ is the set of all differentiable maps of $\mathcal{O}$ to $SO(E^3)$ that take $\rho$ to the same point $f(\rho)$, while having the same differential map at $\rho$ as $df|_\rho : T_\rho \mathcal{O} \to T_{f(\rho)} SO(E^3)$. Thus, the information that defines $j_\rho^1 f$ uniquely is equivalent to the set of coordinates $(\rho^a, x^i, e_j^i, x_a^i, e_{ja}^i)$, where $x_a^i$ and $e_{ja}^i$ represent the matrices of partial derivatives with respect to $\rho^a$ of $x^i$ and $e_j^i$, resp., for any of the representative local embeddings around $\rho$.

---

([1]) To simplify, we have renotated $\nabla$ as $\nabla_\wedge$ in the first step.
([2]) For a more thorough mathematical treatment of the geometry of jet manifolds, one might confer Saunders [**29**]. The paper of Gallisot [**30**] is particularly adapted to the problems of continuum mechanics.

Actually, the constraint on *f* that it be an embedding is unnecessarily global, compared to simply demanding that *f* be an immersion, since jets are purely local objects that are usually indifferent to the behavior of functions outside of some neighborhood of the point in question. Thus, one does not use all of the elements of $J^1(\mathcal{O}, SO(E^3))$, but only the ones for which the ranks of the differential maps are the maximum rank of $p = \dim \mathcal{O}$. Since the differential map $df|_\rho$ can be defined independently of any choice of coordinate charts on $\mathcal{O}$ and $SO(E^3)$, imposing the the maximal rank condition on it defines a level subset of a integer-valued function on $J^1(\mathcal{O}, SO(E^3))$.

The *source projection* $\alpha: J^1(\mathcal{O}, SO(E^3)) \to \mathcal{O}$ simply takes any 1-jet of the form $j^1_\rho f$ to the point $\rho$ in $\mathcal{O}$. Hence, a *section* of the source projection is a differentiable map $s: \mathcal{O} \to J^1(\mathcal{O}, SO(E^3))$ such that the composition of *s*, followed by $\alpha$, gives the identity map on $\mathcal{O}$. Thus, the value of $s(\rho)$ will take the local coordinate form:

$$s(\rho) = (\rho^a, x^i(\rho), e^i_j(\rho), x^i_a(\rho), e^i_{ja}(\rho)).$$

Since we are dealing with differentiable functions now, and not just points in $J^1(\mathcal{O}, SO(E^3))$, we can distinguish between integrable and non-integrable sections of the source projection. In local form, the section *s* is *integrable* iff:

$$s(\rho) = (\rho^a, x^i(\rho), e^i_j(\rho), x^i_{,a}(\rho), e^i_{j,a}(\rho)),$$

in which the commas imply partial derivatives with respect to the parameters $\rho^a$. One also says of such a section that it is the *1-jet prolongation $j^1 f$* of a function $f: \mathcal{O} \to SO(E^3)$, which is then obtained by essentially appending the first partial derivatives of the function. For completeness, we define $J^0(\mathcal{O}, SO(E^3))$ to be simply $\mathcal{O} \times SO(E^3)$, so $j^1 f: J^0(\mathcal{O}, SO(E^3)) \to J^1(\mathcal{O}, SO(E^3))$, where we really mean sections of the source projection, since the differentiation of elements in $J^0(\mathcal{O}, SO(E^3))$ is not defined, anyway.

The *Spencer operator* is then defined to be a map $D: J^1(\mathcal{O}, SO(E^3)) \to \Lambda^1 \mathcal{O} \otimes \mathfrak{X}(SO(E^3))$, that takes the section *s* to a 1-form $Ds$ on $\mathcal{O}$ with values in the vector fields on $SO(E^3)$ that we simply define locally to be:

$$Ds = (dx^i - x^i_a d\rho^a) \otimes \frac{\partial}{\partial x^i} + (de^i_j - e^i_{ja} d\rho^a) \otimes \frac{\partial}{\partial e^i_j}. \tag{3.36}$$

Thus, *Ds* vanishes iff *s* is integrable.

One can also characterize integrability as the vanishing of the pull-backs to $\mathcal{O}$ of the contact 1-forms on $J^1(\mathcal{O}, SO(E^3))$:

$$\vartheta^i = dx^i - x_a^i d\rho^a, \qquad \vartheta_j^i = de_j^i - e_{ja}^i d\rho^a \qquad (3.37)$$

by means of *s*. These pull-backs then take the form:

$$s^* \vartheta^i = (x_{,a}^i - x_a^i) d\rho^a, \qquad s^* \vartheta_j^i = (e_{j,a}^i - e_{ja}^i) d\rho^a, \qquad (3.38)$$

whose vanishing would then imply:

$$x_a^i = \frac{\partial x^i}{\partial \rho^a}, \qquad e_{ja}^i = \frac{\partial e_j^i}{\partial \rho^a}. \qquad (3.39)$$

 *c. The Lie groupoid $J^1(\mathcal{O}; ISO(3))$.* Since our way of deforming an initial state will be by way of the action of $J^1(\mathcal{O}; ISO(3))$ on $J^1(\mathcal{O}; SO(E^3))$, we shall now discuss the structure of that manifold in some detail.  We shall see that although it admits a group structure, that the multiplication of elements is defined only for elements that project to the same point in $\mathcal{O}$.  Thus, one cannot define a Lie group structure on the manifold, but a "Lie groupoid" structure (cf., e.g., [**30, 31**]).

 In general, the way that one prolongs a group of smooth functions with values in a Lie group $G$ to the Lie groupoid $J^1(\mathcal{O}; G)$ is to define the composition of derivative coordinates by differentiating the composition of group elements and omitting the comma. (Here, it is simplest to use linear algebraic groups of matrices.). Thus, the composition of $(\rho, g, g_a)(\rho, g', g_a')$ into $(\rho, g'', g_a'')$ is by way of:

$$g'' = g g', \qquad g_a'' = g_a g' + g g_a'. \qquad (3.40)$$

Note that the composition is defined only when both elements have the value of $\rho$. Thus, in effect, what we are defining is the composition of *sections* of the source projection onto $\mathcal{O}$.

 Under this composition, the (two-sided) identity element for $(\rho, g, g_a) \in J^1(\mathcal{O}; G)$ is $(\rho, e, 0)$, where $e$ is the identity element of $G$. Hence, just as multiplication is not always defined for two arbitrary elements, similarly, different elements will generally have different identity elements, as well.

 From (3.40), the inverse element to $(\rho, g, g_a)$ becomes $(\rho, g, g_a)^{-1} = (\rho, g^{-1}, g_a^{-1})$, where:

$$g_a^{-1} = - g^{-1} g_a g^{-1}. \qquad (3.41)$$

 In particular, every element of $J^1(\mathcal{O}; G)$ has a unique inverse. Thus, $J^1(\mathcal{O}; G)$ does, in fact, have a groupoid structure, and since it is a finite-dimensional differentiable manifold, as well, it becomes a finite-dimensional Lie groupoid of dimension $p + n + pn$.

In the case of $G = ISO(3)$, the composition of $(\rho, a, R, a_a, R_a)(\rho, a', R', a'_a, R'_a)$ into $(\rho, a'', R'', a''_a, R''_a)$ is by way of:

$$a'' = a + Ra', \quad R'' = RR', \quad a''_a = a_a + R_a a' + R a'_a, \quad R''_a = R_a R' + R R'_a. \tag{3.42}$$

The (two-sided) identity element of $(\rho, a, R, a_a, R_a)$ is then $(\rho, 0, I, 0, 0)$ and its inverse is $(\rho, a^{-1}, R^{-1}, a_a^{-1}, R_a^{-1})$, where $R^{-1}$ is the usual inverse matrix to $R$, $a^{-1} = -R^{-1}a$ is the usual inverse to $a$, $R_a^{-1}$ for each $a$ would be the matrix of partial derivatives of $R^{-1}$ with respect to $\rho^a$ for any integrable section of the source projection, and:

$$a_a^{-1} = -R_a^{-1}a + R^{-1}a_a. \tag{3.43}$$

The fact that the elements of $J^1(\mathcal{O}; ISO(3))$ have separate identity elements, depending upon their source projection, is another aspect of Lie groupoids.

Just as the right-invariant vector fields on any Lie group define its Lie algebra, the right-invariant vector fields on sections of the source projection of $J^1(\mathcal{O}; ISO(3))$ define the "Lie algebroid" of $J^1(\mathcal{O}; ISO(3))$. A general vector field on $J^1(\mathcal{O}; ISO(3))$ has the following form relative to the natural frame field of a coordinate chart:

$$\delta X = \delta \rho^a \frac{\partial}{\partial \rho^a} + \delta a^i \frac{\partial}{\partial a^i} + \delta R^i_j \frac{\partial}{\partial R^i_j} + \delta a^i_a \frac{\partial}{\partial a^i_a} + \delta R^i_{ja} \frac{\partial}{\partial R^i_{ja}}, \tag{3.44}$$

Actually, we shall not need the first term in this sum, so we shall omit it from now on.

In order to define the right-invariant vector fields on $ISO(3)$, one must right-translate the second two terms this back to the identity to obtain an element $(\delta \xi^i, \delta I^i_j)$ of the Lie algebra iso(3):

$$\delta \xi^i = \delta a^i - \delta I^i_j a^j, \qquad \delta I^i_j = \delta R^i_k \tilde{R}^k_j. \tag{3.45}$$

We will find it convenient to prolong these component functions by simple partial differentiation with respect to $\rho^a$:

$$\delta \xi^i_a = \delta a^i_a - \delta I^i_{ja} a^j + \delta I^i_j a^j_a, \qquad \delta I^i_{ja} = \delta R^i_{ka} \tilde{R}^k_j + \delta R^i_k \tilde{R}^k_{ja}. \tag{3.46}$$

Thus, the element $(\rho^a, \delta \xi^i, \delta I^i_j, \delta \xi^i_a, \delta I^i_{ja})$ can be regarded as the local coordinates of an element of $J^1(\mathcal{O}; \text{iso}(3))$, and a section of the source projection – i.e., a vector field on an embedded submanifold of $ISO(3)$ – will be integrable iff:

$$\delta \xi^i_a = \partial_a \delta \xi^i, \qquad \delta I^i_{ja} = \partial_a \delta I^i_j. \tag{3.47}$$

*d. The action of $J^1(\mathcal{O}; ISO(3))$ on $J^1(\mathcal{O}; SO(E^3))$.* In order to define the action of $J^1(\mathcal{O}; ISO(3))$ on $J^1(\mathcal{O}; SO(E^3))$, one analogously prolongs the action of $ISO(3)$ on $SO(E^3)$. Recall that one can think of that action as a right action by the inverse element of $ISO(3)$ on points and frames or a left action by an element of $ISO(3)$ on the coordinates of the point and the frame.

In the latter case, if the coordinates of the point $x_0 \in E^3$ are $x_0^i$, and the coordinates of the oriented, orthonormal frame $\mathbf{e}_{i0} = e_{i0}^j \partial_j$ are $e_{i0}^j$ then the action of $(a^i, R_j^i) \in C^\infty(\mathcal{O}; ISO(3))$, in which we have suppressed the functional dependency for the sake of brevity, is:

$$x^i = a^i + R_j^i x_0^j, \qquad e_j^i = R_k^i e_{j0}^k. \tag{3.48}$$

This then differentiates to:

$$x_a^i = a_a^i + R_{ja}^i x_0^j + R_j^i x_{a0}^j, \qquad e_{ja}^i = R_{ka}^i e_{j0}^k + R_k^i e_{ja0}^k. \tag{3.49}$$

Thus, we have defined the action of $(\rho, a, R, a_a, R_a)$ on $(\rho, x_0, \mathbf{e}_{i0}, x_{a0}, \mathbf{e}_{ia0})$ to produce $(\rho, x, \mathbf{e}_i, x_a, \mathbf{e}_{ia})$.

Note that since $(x^0, \mathbf{e}_{i0})$ represents an initial embedding of an object $\mathcal{O}$ in $SO(E^3)$, one cannot assume that its derivatives with respect to $\rho^a$ will vanish.

When one differentiates the action that we just defined, one obtains the fundamental vector field on an embedded object in $SO(E^3)$ that is associated with a section of the source projection of $J^1(\mathcal{O}; \mathfrak{iso}(3))$.

The simplest way to obtain the fundamental vector field is to first apply the operator $\delta$ to the formulas (3.48) and (3.49) like a derivative operator (we shall justify this shortly), while treating the initial coordinates as things that are not affected by the operator:

$$\delta x^i = \delta a^i + \delta R_j^i x_0^j, \qquad \delta e_j^i = \delta R_k^i e_{j0}^k, \tag{3.50}$$

$$\delta x_a^i = \delta a_a^i + \delta R_{ja}^i x_0^j + \delta R_j^i x_{a0}^j, \qquad \delta e_{ja}^i = \delta R_{ka}^i e_{j0}^k + \delta R_k^i e_{ja0}^k. \tag{3.51}$$

This then defines the vector field on $J^1(\mathcal{O}; SO(E^3))$:

$$\delta s = \delta x^i \frac{\partial}{\partial x^i} + \delta e_j^i \frac{\partial}{\partial e_j^i} + \delta x_a^i \frac{\partial}{\partial x_a^i} + \delta e_{ja}^i \frac{\partial}{\partial e_{ja}^i}, \tag{3.52}$$

in which one substitutes the definitions of the components above.

One then right-translates back to the identity by solving for the initial coordinates in terms of the deformed ones and gets:

$$\delta x^i = \delta \xi^i + \delta I_j^i x^j, \qquad \delta e_j^i = \delta I_k^i e_j^k, \tag{3.53}$$

$$\delta x_a^i = \delta \xi_a^i + \delta I_{ja}^i x^j + \delta I_j^i x_a^j, \qquad \delta e_{ja}^i = \delta I_{ka}^i e_j^k + \delta I_k^i e_{ja}^k. \tag{3.54}$$

This then allows one to define the infinitesimal action of $J^1(\mathcal{O}; \mathfrak{iso}(3))$ on $J^1(\mathcal{O}; SO(E^3))$ to produce a vector field on $J^1(\mathcal{O}; SO(E^3))$, namely:

$$\delta s = \delta\xi^i \boldsymbol{\xi}_i + \delta I^i_j \mathcal{I}^j_i + \delta\xi^i_a \boldsymbol{\xi}^a_i + \delta I^i_{ja} \mathcal{I}^{ja}_i, \tag{3.55}$$

in which the anholonomic frame field $\boldsymbol{\xi}_i$, $\mathcal{I}^j_i$, $\boldsymbol{\xi}^a_i$, $\mathcal{I}^{ja}_i$ can be easily related to the natural frame field, although we shall no need for the explicit expressions.

**4. Statics of Cosserat media: Eulerian formulation.** One of the advantages of the present formulation of the kinematics of deformation for Cosserat media is that the static or dynamical state of the medium becomes, in a real sense, "dual" to the kinematical state, or rather, to variations of the kinematical state. The duality is defined relative to the bilinear pairing of dynamical 1-forms and kinematical vector fields that amounts to the definition virtual work.

One finds that all of the relevant constructions can be defined upon either the initial state of the object, which is then the Lagrangian picture, or the deformed state, which is then the Eulerian picture. We shall find that the simplest path to the Cosserat equations is to use the Eulerian picture.

*a. Virtual displacements of the kinematical state.* Now that we have defined a kinematical state of an object in a Cosserat medium to be a section $s: \mathcal{O} \to J^1(\mathcal{O}, SO(E^3))$ of the source projection, it becomes straightforward to define the dynamical state of the object to be the way that it responds to variations of the kinematical state; i.e., virtual displacements. For us, a *variation* or *virtual displacement* $\delta f$ of the embedding $f: \mathcal{O} \to SO(E^3)$, $\rho^a \mapsto (x(\rho), \mathbf{e}_i(\rho))$ will be a vector field on its image $f(\mathcal{O})$. Locally, it will have the component form:

$$\delta f(\rho) = \delta x^i(\rho) \frac{\partial}{\partial x^i} + \delta e^i_j(\rho) \frac{\partial}{\partial e^i_j}. \tag{4.1}$$

The reason that the symbol $\delta$ can be used as if it represented a first-order differential operator is that if one defines a *finite variation* of a differentiable map $f: M \to N$ to be a differentiable one-parameter family $\overline{f}: (-\varepsilon, +\varepsilon) \times M \to N$ of differentiable map from $M$ to $N$ such that $\overline{f}(0, x) = f(x)$ then one can rigorously define the vector field $\delta f$ on the image of $f$ by:

$$\delta f(x) = \left.\frac{\partial \overline{f}(\varepsilon, x)}{\partial \varepsilon}\right|_{\varepsilon=0}. \tag{4.2}$$

Hence, a variation is indeed the result of a first-order, linear, differential operator.

In particular, when $N = \mathbb{R}$ the operator $\delta$ enjoys the following properties:

$$\delta(\alpha f + \beta g) = \alpha \delta f + \beta \delta g, \qquad \delta(fg) = \delta f \, g + f \, \delta g, \qquad \delta(df) = d(\delta f). \quad (4.3)$$

This is useful when dealing with coordinate and component expressions.

Since we are attributing the deformed state $f(\rho)$ to the right action of a smooth function $\chi: \mathcal{O} \to ISO(3)$ on the initial object $f_0(\rho)$, we can describe the virtual displacement $\delta f$ of that object as the restriction of the fundamental vector field $\delta X$ of the action of some smooth function $\delta \chi: \mathcal{O} \to \mathfrak{iso}(3)$, $\rho \mapsto$ on $SO(E^3)$ to the embedded Cosserat object in $SO(E^3)$. Thus, one can get the components of $\delta f$ from:

$$\delta x^i = \delta \xi^i + \delta I^i_j x^j, \qquad \delta e^i_j = \delta I^i_k e^k_j. \quad (4.4)$$

*b. Integrable variations of kinematical states.* We now consider the prolongation of vector fields on $\chi: \mathcal{O} \to SO(E^3)$ to vector fields on $J^1(\mathcal{O}; SO(E^3))$. Now, in order to obtain the prolongation of a vector field $\delta \chi$ on $\chi: \mathcal{O} \to SO(E^3)$ to a vector field on $J^1(\mathcal{O}; SO(E^3))$ – at least, locally – one first differentiates $\delta \chi$:

$$d(\delta \chi) = d(\delta x^i) \otimes \frac{\partial}{\partial x^i} + d(\delta e^i_j) \otimes \frac{\partial}{\partial e^i_j}$$

$$= \partial_a \delta x^i \left( d\rho^a \otimes \frac{\partial}{\partial x^i} \right) + \partial_a \delta e^i_j \left( d\rho^a \otimes \frac{\partial}{\partial e^i_j} \right), \quad (4.5)$$

and then associates the tensor products with natural frame vectors on $J^1(\mathcal{O}; SO(E^3))$, as follows:

$$d\rho^a \otimes \frac{\partial}{\partial x^i} \mapsto \frac{\partial}{\partial x^i_a}, \qquad d\rho^a \otimes \frac{\partial}{\partial e^i_j} \mapsto \frac{\partial}{\partial e^i_{ja}}.$$

We shall call the resulting map $\lambda_V : T^*(\mathcal{O}) \otimes T(SO(E^3)) \to V(J^1(\mathcal{O}; SO(E^3)))$ the *vertical lift* of the tensor product to a vector on $J^1(\mathcal{O}; SO(E^3))$ that is vertical under the contact projection. By contrast, the association of $\delta \chi$ itself with a vector field on $J^1(\mathcal{O}; SO(E^3))$ looks like the identity transformation in local coordinates.

One then assembles the vector $\delta \chi$ and $\lambda_V(d(\delta \chi))$ into the desired prolongation:

$$j_* \delta \chi = \delta \chi + \lambda_V(d(\delta \chi)) = \delta x^i \frac{\partial}{\partial x^i} + \delta e^i_j \frac{\partial}{\partial e^i_j} + \partial_a \delta x^i \frac{\partial}{\partial x^i_a} + \partial_a \delta e^i_j \frac{\partial}{\partial e^i_{ja}}. \quad (4.6)$$

Thus, we are simply saying that:

$$\delta x^i_a = \partial_a \delta x^i, \qquad \delta e^i_{ja} = \partial_a \delta e^i_j. \quad (4.7)$$

We then regard the prolongation map $j_*$ as a map $j_*: T(SO(E^3)) \to T(J^1(\mathcal{O}; SO(E^3)))$, where we are making the aforementioned abbreviation of referring to the ranges of maps from $\mathcal{O}$ to other manifolds, since the operator $j_*$ is meaningful only for those maps, not the points of the manifolds in their ranges. A vector field $\delta X$ on $J^1(\mathcal{O}; SO(E^3))$ is then called *integrable* iff it is the *1-jet prolongation* $j_*\delta\chi$ of a vector field $\delta\chi$ on the image of $\chi$ in $SO(E^3)$.

When one regards the variation $\delta s$ as being that fundamental vector field that comes from the infinitesimal action of a section $\delta X: \mathcal{O} \to J^1(\mathcal{O}; \mathfrak{iso}(3))$ on sections of the source projection of $J^1(\mathcal{O}; SO(E^3))$ as in section **3**.*d*, one sees that $\delta s$ is integrable iff $\delta X$ is integrable, which implies that one must have:

$$\delta \xi^i_a = \partial_a \delta \xi^i \qquad \delta I^i_{ja} = \partial_a \delta I^i_j. \tag{4.8}$$

*c. The dynamical state of an object.* We define the *fundamental 1-form* $\phi$ on $J^1(\mathcal{O}, SO(E^3))$ by:

$$\phi = F_i\, dx^i + M_i^j\, de^i_j + \sigma_i^a\, dx^i_a + \mu_i^{ja}\, de^i_{ja}. \tag{4.9}$$

The components $F_i$ describe the *external forces* that act upon the body, $M_i^j$, the *external moments*, $\sigma_i^a$ represent the *force stresses*, and $\mu_j^{ia}$ represent the *couple stresses*. However, at this point only $F_i$ behaves in the usual way, since $M_i^j$ and $\mu_j^{ia}$ do not have to be anti-symmetric, and $\sigma_i^a$ does not have to be square, much less symmetric. In order to get closer to the usual expressions, one must go to the deformed state.

When one takes into account the action of $J^1(\mathcal{O}; ISO(3))$ on $J^1(\mathcal{O}; SO(E^3))$ one finds that the coordinate differentials take the form of the components of the fundamental vector fields, but with the $\delta$'s replaced with $d$'s:

$$dx^i = \xi^i + \mathcal{I}^i_j x^j, \qquad de^i_j = \mathcal{I}^i_k e^k_j, \tag{4.10}$$

$$dx^i_a = \xi^i_a + \mathcal{I}^i_{ja} x^j + \mathcal{I}^i_j x^j_a, \qquad de^i_{ja} = \mathcal{I}^i_{ka} e^k_j + \mathcal{I}^i_k e^k_{ja}, \tag{4.11}$$

in which we have defined the anholonomic coframe field:

$$\xi^i = dx^i - \mathcal{I}^i_j x^j, \qquad \mathcal{I}^i_j = de^i_k \tilde{e}^k_j, \tag{4.12}$$

$$\xi^i_a = dx^i_a - \mathcal{I}^i_{ja} x^j - \mathcal{I}^i_j x^j_a, \qquad \mathcal{I}^i_{ja} = (de^i_{ka} - \mathcal{I}^i_l e^l_{ka})\tilde{e}^k_j, \tag{4.13}$$

which is, in fact, reciprocal to the anholonomic frame field that we introduced above.

By substituting the previous four expressions for the coordinate differentials into equation (4.9), this gives another form for $\phi$, namely:

$$\phi = \bar{F}_i \xi^i + \bar{M}_{ij} \mathcal{I}^{ij} + \bar{\sigma}_i^a \xi_a^i + \bar{\mu}_{ij}^a \mathcal{I}_a^{ij} \tag{4.14}$$

with:

$$\bar{F}_i = F_i, \qquad\qquad\qquad\qquad \bar{\sigma}_i^a = \sigma_i^a, \tag{4.15}$$
$$\bar{M}_{ij} = M_{[i}^k e_{j]k} + F_{[i} x_{j]} + \sigma_{[i}^a x_{j]a} + \mu_{[i}^{ka} e_{j]ka}, \qquad \bar{\mu}_{ij}^a = \mu_{[i}^{ka} e_{j]k} + \sigma_{[i}^a x_{j]}. \tag{4.16}$$

The anti-symmetrization of the components of $\bar{M}_i^j$ and $\bar{\mu}_i^{ja}$ is necessitated by the anti-symmetry in the *ij*-indices of $\mathcal{I}_j^i$ and $\mathcal{I}_{ja}^i$, as infinitesimal rotations, and it is generally simpler to justify the anti-symmetrization of tensor components when they are both covariant or both contravariant, so we replace $\mathcal{I}_j^i$ and $\mathcal{I}_{ja}^i$ with $\mathcal{I}^{ij}$ and $\mathcal{I}_a^{ij}$, respectively.

Thus, the first two terms of the total moment represent the internal moment and the usual force moment at the point $x^i$ about the origin. The third term is somewhat more unexpected, namely, $\sigma_{[i}^a x_{j]a}$. In the case of point motion, it would be related to the cross product of the momentum with the velocity, and in conventional mechanics, where momentum is always collinear with velocity, it would have to vanish. However, in the mechanics of the Dirac electron, and its simplification in the form of the Weyssenhoff fluid, there is also a "transverse momentum" included in the momentum 1-form, which is not collinear with the covelocity 1-form, and this new contribution could actually play a physically meaningful role. Finally, the last term is also foreign to the usual discussions of classical mechanics, although it will drop out in the final equations.

*d. Constitutive laws.* The constitutive laws for the Cosserat medium are essentially included in the functional dependency of the components of $\phi$ on the kinematical state, so they take the form:

$$F_i = F_i(\rho^a, x^i, e_j^i, x_a^i, e_{ja}^i),$$
$$M_i^j = M_i^j(\rho^a, x^i, e_j^i, x_a^i, e_{ja}^i),$$
$$\sigma_i^a = \sigma_i^a(\rho^a, x^i, e_j^i, x_a^i, e_{ja}^i),$$
$$\mu_i^{ja} = \mu_i^{ja}(\rho^a, x^i, e_j^i, x_a^i, e_{ja}^i).$$

Of course, most practical constitutive laws take simpler forms than these expressions. Moreover, most practical constitutive laws only pertain to the case of infinitesimal displacements, so the distinction between the Lagrangian and Eulerian pictures disappears.

A discussion of constitutive laws for Cosserat media, including the example of a linear, isotropic, holonomic medium, can be found in [**17**, **18**, **25**, **26**].

*e. Virtual work.* The response of the object *f* to a virtual displacement $\delta s$ of its kinematical state *s* will be an increment of *virtual work*:

$$\delta W = \phi[\delta s]. \tag{4.17}$$

According to d'Alembert's principle, the state *s* is in equilibrium iff the virtual work done by all virtual displacements (usually, with certain restrictions) vanishes.

In the general picture, one will have:

$$\delta W = \phi(\delta s) = F_i\, \delta x^i + M_i^{\,j} \delta e_j^i + \sigma_i^a \delta x_a^i + \mu_i^{ja} \delta e_{ja}^i. \tag{4.18}$$

If the virtual displacement is integrable – so $\delta s = j_* \delta f$ – then this becomes:

$$\delta W = F_i\, \delta x^i + M_i^{\,j} \delta e_j^i + \sigma_i^a (\partial_a \delta x^i) + \mu_i^{ja}(\partial_a \delta e_j^i). \tag{4.19}$$

Applying the product rule for differentiation, one finds:

$$\delta W = (F_i - \partial_a \sigma_i^a)\delta x^i + (M_i^{\,j} - \partial_a \mu_i^{ja})\delta e_j^i + \partial_a(\sigma_i^a \delta x^i + \mu_i^{ja} \delta e_j^i). \tag{4.20}$$

In order to account for the last term in this expression, one must then compute the total virtual that is done by $\delta W$, which is obtained by pulling down the function $\delta W$ on $J^1(\mathcal{O}, SO(E^3))$ to functions on $\mathcal{O}$ by means of the kinematical state $s_L$ and integrating it over $\mathcal{O}$ using the volume element:

$$V_p = d\rho^1 \wedge \ldots \wedge d\rho^p.$$

If $\mathcal{O}$ has a boundary $\partial \mathcal{O}$ then the total virtual work done by the virtual displacement will take the form:

$$W[\delta s] = \int_{\mathcal{O}} (j^*\phi_i \delta x^i + j^*\phi_i^{\,j} \delta e_j^i) V_p + \int_{\partial \mathcal{O}} (\sigma_i^a \delta x^i + \mu_i^{ja} \delta e_j^i) \#\partial_a, \tag{4.21}$$

in which:

$$j^*\phi_i = F_i - \partial_a \sigma_i^a, \qquad\qquad j^*\phi_i^{\,j} = M_i^{\,j} - \partial_a \mu_i^{ja}, \tag{4.22}$$

and:

$$\#\partial_a = i_{\partial_a} V_p = \frac{1}{(p-1)!} \varepsilon_{i_1 \cdots a \cdots i_p} d\rho^{i_1} \wedge \cdots \wedge \widehat{d\rho^a} \wedge \cdots \wedge d\rho^{i_p} \tag{4.23}$$

gives the volume element that is induced on the boundary.

Our choice of notation for $j^*$ is based in the fact that it is "essentially" the adjoint of the prolongation operator $j_*$ under the bilinear pairing of 1-forms on $J^1(\mathcal{O}; SO(E^3))$ with vector fields on that manifold:

$$<j^*\phi, \delta f> = <\phi, j_* \delta f> - d<\#\Pi, \delta f>, \tag{4.24}$$

in which:

$$j^*\phi = (F_i - \partial_a \sigma_i^a) dx^i + (M_i^{\,j} - \partial_a \mu_i^{ja}) de_j^i, \tag{4.25}$$

and we have defined the 1-form $\Pi$ on $SO(E^3)$ that takes its values in the vector fields on $\partial \mathcal{O}$:

$$\Pi = (\sigma_i^a dx^i + \mu_i^{ja} de_j^i) \otimes \partial_a. \tag{4.26}$$

Since this 1-form is defined on the boundary of the object, the components $\sigma_i^a$ define the surface tensions while the $\mu_i^{ja}$ define the surface couple-stresses.

In the Eulerian formulation, one uses the fundamental vector field $\delta s$ on the deformed state, as defined by (3.55), in order to obtain the virtual work that is done by such a virtual displacement in the form:

$$\delta W = \phi(\delta s) = F_i \, \delta \xi^i + \bar{M}_{ij} \delta I^{ij} + \sigma_i^a \delta \xi_a^i + \bar{\mu}_{ij}^a \delta I_a^{ij}, \tag{4.27}$$

with the definitions that were given above.

When the variation is integrable – so $\delta \xi_a^i = \partial_a \delta \xi^i$, $\delta I_{ja}^i = \partial_a \delta I_j^i$ – $\delta W$ takes the form:

$$\delta W = (F_i - \partial_a \sigma_i^a) \delta \xi^i + (\bar{M}_{ij} - \partial_a \bar{\mu}_{ij}^a) \delta I^{ij} + \partial_a (\sigma_i^a \delta \xi^i + \bar{\mu}_{ij}^a \delta I^{ij}), \tag{4.28}$$

so, in this case, one can set:

$$j^* \phi_i = F_i - \partial_a \sigma_i^a, \qquad j^* \phi_i^j = \bar{M}_i^j - \partial_a \bar{\mu}_i^{ja}, \qquad \Pi = (\sigma_i^a \delta \xi^i + \bar{\mu}_i^{ja} \delta I_j^i) \otimes \partial_a. \tag{4.29}$$

Hence, the expression that relates to infinitesimal displacements is unaffected, while the expressions that relate to infinitesimal rotations must be altered slightly, along with the nature of the surface stresses.

*f. Euclidian fundamental 1-forms.* A key step in obtaining the Cosserat equations is to restrict oneself to fundamental 1-forms that are *Euclidian*, in the sense that the virtual work that is done by the prolongation of any rigid motion must be zero. Intuitively, one expects this to imply that both the external forces and moments must vanish, but when one does the calculations, one finds that only half of that intuition is true and the condition on the moments amounts to a statement that the moments must be *internal*; i.e., due to the force and couple stresses.

In order to obtain the condition on $\phi$ that corresponds to the stated constraint, it simplest to use the expression (4.27) when one assumes that the components $\delta \xi^i$ and $\delta I_j^i$ are constant functions, so $\delta \xi_a^i = \partial_a \delta \xi^i = 0$, $\delta I_{ja}^i = \partial_a \delta I_j^i = 0$. If $\delta W_E$ vanishes for every such virtual displacement then one must conclude that that the coefficients of $\delta \xi^i$ and $\delta I_j^i$ must vanish accordingly, which means that:

$$F_i = 0, \qquad M_{[i}^k e_{j]k} = -\sigma_{[i}^a x_{j]a} - \mu_{[i}^{ka} e_{j]ka}. \tag{4.30}$$

It is the condition on the moments that represents the key to understanding the Cosserat equations, because it amounts to the idea that the internal force and couple stresses give rise to an internal couple when they are coupled to the gradient of the deformation.

When one includes the results of (4.30) a Euclidian fundamental 1-form can be expressed in the general form:

$$\phi = \sigma_i^a \xi_a^i + (\bar{\mu}_{ij}^a + \sigma_{[i}^a x_{j]}) \mathcal{I}_a^{ij}. \tag{4.31}$$

*g. The equilibrium equations.* When one introduces the usual boundary conditions on the virtual displacements – viz., that they vanish for a fixed boundary or are transverse to Π for a free one – one sees that d'Alembert's principle that the virtual work must vanish for all such virtual displacements implies that:

$$j^* \phi = 0. \tag{4.32}$$

When one starts with $\delta W$ in the general form (4.18), this condition would imply the local equations:

$$F_i = \partial_a \sigma_i^a, \qquad \bar{M}_{ij} = \partial_a \bar{\mu}_{ij}^a, \tag{4.33}$$

which we then rewrite in the form:

$$F_i = \partial_a \sigma_i^a, \qquad M_{[ij]} + F_{[i} x_{j]} + \sigma_{[i}^a x_{j]a} = \partial_a \bar{\mu}_{ij}^a + \partial_a \sigma_{[i}^a x_{j]} + \sigma_{[i}^a \partial_a x_{j]}, \tag{4.34}$$

and after taking the first of these sets of equations into account, and assuming that the initial state is integrable (i.e., $x_a^i = \partial_a x^i$), one gets:

$$F_i = \partial_a \sigma_i^a, \qquad M_{[ij]} = \partial_a \mu_{ij}^a. \tag{4.35}$$

These are the equilibrium equations that one might expect, but they are not the Cosserat equations. In order to obtain the usual Cosserat equations, however, one must first start with a Euclidian fundamental 1-form, and then make some further restrictions and conversions.

One starts with the virtual work that is done by a virtual displacement that takes the form of a fundamental vector field $\delta s$ on the deformed state when the fundamental 1-form is Euclidian, as it is described in (4.31), namely:

$$\delta W_E = \sigma_i^a \delta \xi_a^i + (\bar{\mu}_i^{ja} + \sigma_{[i}^a x^{j]}) \delta I_{ja}^i. \tag{4.36}$$

The integrability condition on $\delta s$ then takes the form:

$$\delta \xi_a^i = \partial_a \delta \xi^i, \qquad \delta I_{ja}^i = \partial_a \delta I_j^i. \tag{4.37}$$

With this, and the usual variational argument, one then derives the set of equilibrium equations:

$$0 = \partial_a \sigma_i^a, \qquad 0 = \partial_a \bar{\mu}_{ij}^a + \sigma_{[i}^a x_{j]a}. \tag{4.38}$$

These are almost the usual form of the Cosserat equations, except that one has to convert from partial derivatives with respect to the object parameters to partial derivatives with respect to the deformed coordinates, if that is possible.

First, one notes that the equations (4.38) are valid for any dimension of object $\mathcal{O}$ up to the dimension of the space $E^n$ in which it is embedded. Thus, no assumption of invertibility is being made about the matrix $\partial_a x^i = \partial x^i / \partial \rho^a$. If one specializes the discussion to three-dimensional objects in three-dimensional space (so $a, i = 1, 2, 3$) then since we are assuming that the initial state comes from an embedding, this matrix will have to be invertible, and we denote its inverse by $\tilde{x}_i^a$.

By the chain rule, one will then have:

$$\frac{\partial}{\partial x^i} = \frac{\partial \rho^a}{\partial x^i} \frac{\partial}{\partial \rho^a} = \tilde{x}_i^a \frac{\partial}{\partial \rho^a}. \qquad (4.39)$$

The three-dimensional Cosserat equations then take the form:

$$0 = \partial_j \sigma_i^j, \qquad 0 = \partial_k \bar{\mu}_{ij}^k + \sigma_{[ij]}. \qquad (4.40)$$

In these equations, the partial derivatives are now taken with respect to the deformed coordinates $x^i$.

In most of the more recent (i.e., post-1950's) literature on Cosserat media, equations (4.40) are the starting point. However, one sees that in the original Cosserat book other dimensions were treated besides three, so the excess volume of this study up to now is partially justified by that fact, along with the deeper fact that here we are starting in the geometry of the Lie group of rigid motions and deriving the Cosserat equations from that more primitive point, rather than assuming them as given.

**5. Discussion.** A natural first direction in which to expand the analysis presented here into the realm of dynamics. In effect, the basic machinery of the dynamics of Cosserat media is already present in the aforementioned treatment of Cosserat statics. The main alterations start with singling out one of the parameters $\rho^a$ as being a time parameter instead of a spatial one. Thus, one finds that the statics of a string and the dynamics of a point are closely related, as are the statics of a surface and the dynamics of a string, etc.

However, in addition to singling out which of the generalized velocities $\xi_a^i$ and $\omega_{ja}^i$ are true velocities, one must also eventually address the fact that in neither non-relativistic nor relativistic mechanics does the time coordinate truly play the same role as the spatial coordinates, since even in relativity one must treat it differently in the signature type of the metric. Thus, a more thorough treatment of Cosserat dynamics would again require a lengthy discussion in its own right, so we will defer that to some later study.

One eventually comes to see that only so much of the methods that we developed above are specific to three-dimensional spaces or the rigid motions, in particular. More to the point, as long as one is dealing with a group *G* that is a semi-direct product $\mathbb{R}^n \times_s$ *GL*(*n*) of a translation group $\mathbb{R}^n$ with a subgroup of *GL*(*n*), which means that it will be a subgroup of *A*(*n*), most of the basic definitions – especially, the fundamental sequence – will remain intact. Thus, one can think in terms of "mechanics with values in a semi-direct product," just as translational and rotational mechanics can be generalized to mechanics with values in a Lie group.

The first physically interesting possibility that presents itself is four-dimensional Minkowski space, so $\mathbb{R}^3$ gets replaced with $\mathbb{R}^4$, and the rigid rotations *SO*(3) get replaced by the orientation-preserving Lorentz transformations of *SO*(1, 3). The resulting semi-direct produce group $\mathbb{R}^4 \times_s$ *SO*(1, 3) is referred to as the *Poincaré group.* As we will see in a later article, the resulting equations of motion include the well-known Weyssenhoff equations as a special case. The relativistic, spinning fluid that they describe is a simplification of the one that is described by tensorial form of the Dirac equation, which describes the wave function of the electron, among other things. Of course, the conventional interpretation for the density that one derives from the Dirac wave function is that of a probability density function, not a mass density function, which is more definitive of the "hydrodynamical" interpretation.

In addition to the issues that arise in relativistic hydrodynamics and the hydrodynamical models for quantum physics, one can also consider relativistic continuum mechanics more generally. For instance the relativistic theory of elasticity, although not as well developed as relativistic hydrodynamics, is nonetheless still potentially useful in the context of neutron stars and black holes, and possibly still measurable at the levels of stresses found in more conventional celestial bodies.

Ultimately, one sees that a fundamental problem in mechanics is how one is supposed to extend the concept of translation from points to extended deformable objects when the concept of a displacement vector field is as general as all diffeomorphisms and the issue of separating out a constant displacement for an extended body is as ambiguous as the number of points in the object, if not the space that it lives in.

# References

## Literature leading up to Cosserat book

## Literature on Cosserat media and couple stresses after Cosserat